\documentclass[aps,prb,twocolumn]{revtex4}
\usepackage{amsmath, amsthm, amssymb,graphicx}
\begin{document}
\title{Spontaneous Symmetry Breaking in Quantum Mechanics}
\author{Jasper van Wezel and Jeroen van den Brink}
    
\affiliation{
Institute-Lorentz for Theoretical Physics,  Universiteit  Leiden,
P.O. Box 9506, 2300 RA Leiden, The Netherlands}
\date{\today}

\begin{abstract}
We present a clear and mathematically simple procedure explaining spontaneous symmetry breaking in quantum mechanical systems. The procedure is applicable to a wide range of models and can be easily used to explain the existence of a symmetry broken state in crystals, antiferromagnets and even superconductors. It has the advantage that it automatically brings to the fore the main players in spontaneous symmetry breaking: the symmetry breaking field, the thermodynamic limit, and the global excitations of the thin spectrum.
\end{abstract}

\maketitle

\section{Introduction}
In quantum mechanics symmetry has a much more powerful role than in classical mechanics. Translational invariance in a classical setup causes momentum to be conserved; in quantum mechanics it immediately implies that all eigenstates of the Hamiltonian are spread out with equal amplitude over all of space. Using this line of reasoning, it could be argued that since a chair is built up out of many microscopic particles, which all obey the rules of quantum mechanics, the chair as a whole should in fact also respect the symmetry of its Hamiltonian and be spread out across all of space. Clearly this is not the physically realized situation. The way out of the paradox is the spontaneous symmetry breaking of the collective system. The description of spontaneous symmetry breaking in macroscopic objects which are constructed from microscopic, quantum mechanical constituents, is one of the highlights of modern condensed matter theory.\cite{Anderson72,Anderson52,Lieb62,Kaiser89,Kaplan90} It is used to explain the classicality of macroscopic systems ranging from crystals and antiferromagnets all the way to superconductors.\endnote{Notice that ferromagnetism is explicitly \emph{not} included in this list. The ferromagnet has a large number of possible exact groundstates which are all precisely degenerate, and which all have a finite magnetization. The singling out of one of these eigenstates is in a sense more like classical symmetry breaking than like the quantum symmetry breaking discussed here. The quantum symmetry breaking causes a state which is not an eigenstate of the Hamiltonian to be realized, and thus goes much further than only singling out one particular eigenstate.}

The general idea behind spontaneous symmetry breaking is easily formulated: as a collection of quantum mechanical particles grows larger, the object as a whole becomes ever more unstable against small perturbations. In the end even an infinitesimal perturbation is enough to cause the collective system to break the underlying symmetry of the Hamiltonian. The fact that the symmetry breaking can happen spontaneously is then signaled by a set of non-commuting limits: In the complete absence of perturbations even a macroscopic system should conform to the symmetry of the Hamiltonian. If on the other hand the system is allowed to grow to macroscopic size in the presence of even just an infinitesimal perturbation, then it will be able to break the symmetry and end up in a classical state. This clear intuitive picture of spontaneous symmetry breaking is unfortunately not always easy to demonstrate in an equally clear mathematical description of the process. In this paper we present a simple mathematical procedure that can be applied to the spontaneous breaking of any continuous symmetry and that naturally emphasized the roles of the key players in this process. The procedure is described by considering the example of a quantum harmonic crystal which spontaneously breaks translational symmetry. However, all of the methods, i.e. bosonization, using the Bogoliubov transformation to identify the thin spectrum of states involved in spontaneous symmetry breaking, introducing a symmetry breaking field in the collective dynamics and finally considering a non-commuting order of limits, can be easily transferred to other cases as well.\cite{vanWezel05,vanWezel06,vanWezel06:2}

\section{The Harmonic Crystal}
As the most basic example of spontaneous symmetry breaking, we consider how translational symmetry is broken in a crystalline lattice.\cite{vanWezel06} Consider the textbook example of a harmonic crystal, with the Hamiltonian
\begin{eqnarray}
H=\sum_{j} \frac{{ p}^2_j}{2 m} + \frac{\kappa}{2} \sum_{j}  \left( {
  x}_j - { x}_{j+1} \right)^2,
\label{eq:Xtal}
\end{eqnarray}
where $j$ labels all $N$ atoms in the lattice, which have mass $m$, momentum ${ p}_j$ and position ${ x}_j$. We consider here only a one-dimensional chain of atoms, but all of the following can be straightforwardly generalized to higher dimensions as well. $\kappa$ parameterizes a harmonic potential between neighboring atoms; the results on spontaneous symmetry breaking that follow however, are equally valid for an-harmonic potentials.\cite{vanWezel06}

In the standard treatment of the harmonic oscillator one uses a Fourier transformation of the Hamiltonian to be able to identify its eigenstates. We follow a slightly longer route by introducing boson (phonon) operators from the outset, and diagonalizing them using a so called Bogoliubov transformation.\cite{Martin:Bogoliubov} This has the advantage that it naturally brings to the fore the thin spectrum of the quantum crystal, and that it enables us to keep track of the center of mass motion of the crystal as a whole.\cite{Martin:thin} 
The momentum and position operators can be expressed in terms of bosonic operators as
\begin{eqnarray}{p}_j = i C \sqrt{\frac{\hbar}{2}} (b^{\dagger}_{j} - b^{\phantom\dagger}_j) ;
~~{x_j} = \frac{1}{C}\sqrt {\frac{\hbar}{2}} (b^{\dagger}_{j} + b^{\phantom\dagger}_j), 
\end{eqnarray}
so that the commutation relation  $[{x_j}, {p_{j'}}] = i \hbar \delta_{j,j'}$
is fulfilled. We choose $C^2 = \sqrt{2m \kappa}$ so that the Hamiltonian reduces to
\begin{eqnarray}
H &=& \frac{\hbar}{4} \sqrt \frac{2 \kappa}{m} \sum_{j} \left[ 2 
(b^{\dagger}_{j} b^{\phantom\dagger}_j + b^{\phantom\dagger}_j b^{\dagger}_{j}) \right. \nonumber \\
& & \hspace{50pt} \left. - (b^{\dagger}_{j} + b^{\phantom\dagger}_j) (b^{\dagger}_{j+1}+b^{\phantom\dagger}_{j+1}) \right],
\end{eqnarray}
and after a Fourier transformation 
\begin{eqnarray} H &=&  
\hbar \sqrt{\frac{\kappa}{2m}} \sum_k \left[ A_k b^{\dagger}_{k} b^{\phantom\dagger}_k + \frac{B_k}{2} 
(b^{\dagger}_k b^{\dagger}_{-k} + b^{\phantom\dagger}_k b^{\phantom\dagger}_{-k}) +1 \right], \nonumber 
\end{eqnarray}
where $A_k =2 - \cos \left( ka \right)$, $B_{k} = - \cos \left( ka \right)$
and $a$ is the lattice constant. This Hamiltonian is still not diagonal, since
the terms $b^{\dagger}_k b^{\dagger}_{-k}$ and $b^{\phantom\dagger}_{k}
b^{\phantom\dagger}_{-k}$ create and annihilate two bosons at the same
time. We get rid of these terms by introducing transformed bosons 
$\beta^{\phantom\dagger}_k=\cosh(u_k) b^{\phantom\dagger}_{-k} + \sinh(u_k)
b^{\dagger}_k$, and choosing $u_k$ such that the resulting Hamiltonian will be diagonal. After this Bogoliubov transformation, the Hamiltonian in terms of transformed bosons is given by
\begin{eqnarray}
H &=& \hbar \sqrt \frac{\kappa}{m} \sum_k \left[ 2 \sin |ka/2| \left( \beta^{\dagger}_{k}
\beta^{\phantom\dagger}_{k} + \frac{1}{2} \right) \right. \nonumber \\
&+& \left. \frac{1}{4} \sqrt {2}\cos\left(ka\right) \right]\nonumber \\ 
&=& 2 \hbar {\sqrt \frac{\kappa}{m}} \sum_k \sin |ka/2| \left[ n_k +\frac{1}{2} \right],
\label{eq:Hbeta}
\end{eqnarray}
since  $\sum_k \cos k = \frac{N}{2 \pi} \int^{\pi}_{- \pi} d k \cos
k = 0$. 

\section{The Thin Spectrum}
The final form of the Hamiltonian in terms of phonon operators of course coincides with the standard text book result, but the use of the Bogoliubov transformation to get here has the advantage that it draws attention to a rather subtle point. When $k \to 0$ the excitation energy $\omega_k \to 0$ and the two parameters in the Bogoliubov transformation diverge ($\sinh(u_k) \to \infty$ and $\cosh(u_k) \to \infty$). Precisely at $k = 0$ the canonical transformation is thus no longer well defined. This means that we should really treat the $k=0$ part of the Hamiltonian~\eqref{eq:Xtal} separately from the rest.\cite{Martin:thin} The excitations with $k=0$ are the ones that describe the collective dynamics of the quantum crystal as a whole, and therefore they are also precisely the states that are involved in the collective symmetry breaking. The $k=0$ part of the Hamiltonian, written again in terms of the original operators, is given by
\begin{eqnarray}
H_{coll}= \frac{p^2_{tot}}{2 N m} + \text{constant},
\label{Hcoll}
\end{eqnarray}
where ${p}_{tot} \equiv \sum_j {p}_j=\sqrt{N} {p}_{{\bf k}=0}$ is
the total momentum of the entire system, or equivalently, its center of mass
momentum. It can easily be checked that this part of the Hamiltonian, which describes the external dynamics of the crystal as a whole, in fact commutes with the rest of the Hamiltonian, which describes the internal dynamics of the phonon modes inside the crystal. We therefore focus on this collective part of the Hamiltonian from now on, and disregard the phonon spectrum given by~\eqref{eq:Hbeta}.

The eigenstates of the collective Hamiltonian $H_{coll}$ are very low in energy: their excitation energies scale with $1/N$, where $N$ is the number of atoms in the crystal. In the thermodynamic limit all of these states thus become nearly degenerate. It is because of this property that in the thermodynamic limit a combination of these states which breaks the symmetry of the Hamiltonian can be spontaneously formed. On the other hand these collective eigenstates are so few in number and of such low energy that their contribution to the free energy completely disappears in the thermodynamic limit. This can be easily understood by looking at their contribution to the partition function
\begin{eqnarray}
Z_{thin} &=& \sum e^{-\beta H_{coll}} \propto \sqrt{N} \nonumber \\
F_{thin} &=& -T \ln \left( Z_{thin} \right) \propto \ln \left( N \right) .
\end{eqnarray}
The free energy of the total system is an extensive quantity, so that $F_{thin}/F_{tot} \propto \ln (N) / N$ disappears in the limit $N \to \infty$. The states of this part of the spectrum are thus invisible in thermodynamically measurable quantities such as for instance the specific heat of macroscopic crystals, and it is consequently called the \emph{thin spectrum} of the quantum crystal.

To see how the states in the thin spectrum can conspire to break the translational symmetry, we need to add a small symmetry breaking field to the Hamiltonian:
\begin{eqnarray}
H_{coll}^{SB}= \frac{p^2_{tot}}{2 N m} + \frac{B}{2} {x}_{tot}^2.
\label{eq:HcollSB}
\end{eqnarray}
Here the symmetry breaking field $B$ is introduced as a mathematical tool, and need not really exist. In fact, we will send the value of $B$ to zero at the end of the calculation. The Hamiltonian~\eqref{eq:HcollSB} is the standard form of the Hamiltonian for a quantum harmonic oscillator, and its eigenstates are well known. The groundstate wavefunction can be written as
\begin{eqnarray}
\psi_0 (x_{tot}) =\left( \frac{m \omega N}{\pi\hbar} \right)^{1/4}
e^{-\frac{m\omega N}{2 \hbar} x_{tot}^2},
\end{eqnarray}
with $\omega=\sqrt{\frac{B}{m N}}$. This groundstate is a wavepacket of the total momentum states that make up the thin spectrum. Apart from the groundstate configuration there are also collective eigenstates that are described by the excitations of the harmonic oscillator equation~\eqref{eq:HcollSB}. These excitations describe the collective motion of the crystal as a whole. As $N$ grows larger the groundstate wavepacket becomes more and more localized at the position $x_{tot}=0$, until it is completely localized as $N \to \infty$. That this localization can in fact occur spontaneously, without the existence of a physical symmetry breaking field $B$ can be seen by considering the non commuting limits
\begin{eqnarray}
\lim_{N\to \infty} \lim_{B\to 0} \psi_0 (x_{tot}) &=& \text{const} \nonumber \\
\lim_{B\to 0} \lim_{N\to \infty} \psi_0 (x_{tot}) &=& \delta_{ x_{tot},0}.
\end{eqnarray}
If we do not include any symmetry breaking field at all then the crystal is always completely delocalized, and respects the symmetry of the Hamiltonian. If on the other hand we do allow for a symmetry breaking field, then it turns out that in the limit of having infinitely many constituent particles, an infinitesimally small symmetry breaking field is enough to completely localize the crystal in a single position. This mathematical instability clearly implies that in the thermodynamic limit the symmetry breaking in fact happens spontaneously.

To see in a more rigorous manner whether or not the crystal as a whole is localized, we should look at the spatial fluctuations of the crystal: $\left< x_{tot}^2 \right>$. In itself however, the size of these fluctuations is meaningless. The fluctuations become meaningful only if they are compared to the size of the crystal itself. Because the size of the crystal is directly proportional to the number of particles in the system, the correct orderparameter to look at in this case is $\left< x_{tot}^2 \right> /N$. This orderparameter has a non-commuting order of limits as $N \to \infty$
\begin{eqnarray}
\lim_{N\to \infty} \lim_{B\to 0} \left< x_{tot}^2 \right> /N &=& \infty \nonumber \\
\lim_{B\to 0} \lim_{N\to \infty} \left< x_{tot}^2 \right> /N &=& 0,
\end{eqnarray}
which again signals the spontaneous localization of the crystal as a whole. 

\section{Subtleties}
In the derivation of the spontaneous symmetry breaking of a harmonic crystal we have been somewhat sloppy in the definition of the symmetry breaking field. After all, the collective model of equation~\eqref{Hcoll} was only the $k=0$ part of the full blown Hamiltonian~\eqref{eq:Xtal}, but we did not consider the symmetry breaking field to be only the $k=0$ part of some other field acting on all atoms individually. It would therefore be better to start with a microscopic model which already includes a symmetry breaking field, like for example
\begin{eqnarray}
H^{SB} &=& \sum_{j} \left[ \frac{{ p}^2_j}{2 m} + \frac{\kappa}{2} \left( {
  x}_j - { x}_{j+1} \right)^2  \right. \nonumber \\
  & & \left. \phantom{\frac{{ p}^2_j}{2 m} +} + B \left(1-\cos \left( x_j \right) \right) \right] \nonumber \\
\Rightarrow H_{coll}^{SB} &\simeq& \frac{p^2_{tot}}{2 N m} + \frac{B}{2N} {x}_{tot}^2.
\label{eq:HB}
\end{eqnarray}
where in the last line we again consider only the $k=0$ part of the Hamiltonian, and we have expanded the cosine to quadratic order. The fact that the symmetry breaking field now scales as $1/N$ is a direct consequence of our definition of the microscopic symmetry breaking field. In fact the factor $1/N$ cannot be avoided if we insist that the microscopic Hamiltonian be extensive. This may seem to imply an end to the localization of the total wavefunction $\psi_0 \left( x_{tot} \right)$, but in fact spontaneous symmetry breaking is still possible as long as we consider the correct orderparameter. Even though the wavefunction itself does not reduce to a delta function anymore, the spatial fluctuations of the crystal as compared to its size do still become negligible in the thermodynamic limit if an infinitesimal symmetry breaking field is included:
\begin{eqnarray}
\lim_{N\to \infty} \lim_{B\to 0} \left< x_{tot}^2 \right> /N &=& \infty \nonumber \\
\lim_{B\to 0} \lim_{N\to \infty} \left< x_{tot}^2 \right> /N &=& 0.
\end{eqnarray}
Once again the disappearance of fluctuations in the thermodynamic limit signals the spontaneous localization of the crystal as a whole.

For a clear view on the essential ingredients of spontaneous symmetry breaking, this digression into extensivity and a correct choice for the symmetry breaking field seems unnecessary and therefore we have chosen to ignore these subtleties in our main treatment of quantum mechanical spontaneous symmetry breaking. In the application of this procedure to other systems, such as antiferromagnets and superconductors, these issues don't come up because one is forced to consider extensive models from the outset. On the other hand, in those cases the mathematics of diagonalizing the collective Hamiltonian is a bit more involved.\cite{vanWezel06,vanWezel06:2}

\section{Discussion}
We have presented in this paper a simple way of mathematically underpinning the explanation of the effect of spontaneous symmetry breaking in quantum mechanical systems.
The procedure starts out with the bosonization of the microscopic Hamiltonian. The quadratic part of the bosonized Hamiltonian can in principle be diagonalized using a Bogoliubov transformation, but in doing so one finds that there are some modes for which the transformation is ill-defined. It can be shown that these singular modes are precisely the ones describing the dynamics of the system as a whole (as opposed to the dynamics of constituent particles within the system). These collective excitations should be treated separately from all other modes, and together they define the collective part of the Hamiltonian of the system. The eigenstates of this collective Hamiltonian which scale as $1/N$ form the so called thin spectrum, and it is a combination of these states that will make up the symmetry broken wavefunction in the end. As a mathematical tool necessary to be able to see the symmetry breaking explicitly, we introduce the symmetry breaking field $B$. If we then look at the new groundstate wavefunction, or at a suitably defined order parameter for the system, then we see that in the thermodynamic limit even an infinitesimally small field $B$ is enough to completely break the symmetry of the underlying Hamiltonian. It is thus argued that in the limit of $N\to \infty$ the symmetry breaking can in fact happen spontaneously.

The method as presented here can be easily adopted to describe rotors, antiferromagnets, and even superconductors and should in principle be applicable to all quantum mechanical systems which spontaneously break some continuous symmetry.

\end{document}